\documentclass[twocolumn,english,pra,showpacs,preprintnumbers,amsmath,amssymb]{revtex4-1}
\usepackage[latin9]{inputenc}
\usepackage{bm}
\usepackage{amsmath}
\usepackage{amsthm}
\usepackage{amssymb}
\usepackage{stmaryrd}
\usepackage{graphicx}

\makeatletter

\providecommand{\tabularnewline}{\\}

\usepackage{color}

\providecommand{\tabularnewline}{\\}

\usepackage{epsfig}\usepackage{subfigure}\usepackage{dcolumn}\usepackage{mathrsfs}\usepackage{pifont}\usepackage{amsthm}
\usepackage{bm}\usepackage{color}\usepackage{latexsym}\usepackage{amsfonts}

\usepackage{babel}

\makeatother

\usepackage{babel}
\begin{document}
\title{Measurement of tunnel coupling in a Si double quantum dot based on charge sensing}
\author{Xinyu Zhao}
\author{Xuedong Hu}
\email{xhu@buffalo.edu}

\affiliation{Department of Physics, University at Buffalo, SUNY, Buffalo, New York
14260, USA}
\begin{abstract}
In Si quantum dots, valley degree of freedom, in particular the generally small valley splitting and the dot-dependent valley-orbit phase, adds complexities to the low-energy electron dynamics and the associated spin qubit manipulation.  Here we propose a four-level model to extract tunnel coupling information for a Si double quantum dot (DQD).  This scheme is based on a charge sensing measurement on the ground state as proposed in the widely used protocol for a GaAs double dot {[}DiCarlo et. al.,
PRL 92. 226801{]}.  Our theory can help determine both intra- and inter-valley tunnel coupling with high accuracy, and is robust against system parameters such as valley splittings in the individual quantum dots.
\end{abstract}
%\pacs{73.63-b, 72.25.Rb, 03.67.Hk.}
\maketitle

\section*{Introduction}

Tunnel coupling is an essential element in coherent manipulation of electron qubits in semiconductor quantum dots (QDs) \citep{zwanenburg2013RMP, hanson2007RMP, gullans2020PRB, mills2019NC, petta2010Science, otxoa2019PRB, kervinen2019PRL, penthorn2019npjQI, joecker2019PRB, zhang2019NSR, cota2018JPCM, martins2017PRL, schoenfield2017NatComm, chen2017PRB, culcer2010PRB, culcer2010PRB2, li2010PRB, li2017PRA}.  It allows single-qubit operations on a charge qubit, and exchange gates for spin qubits \citep{shi2012PRL, koh2012PRL, kim2014Nature, cao2016PRL,ferraro2014QIP, taylor2005NatPhys, takeda2020PRL}.  Interdot shuttling is also crucial for information transfer on chip \citep{ginzel2020PRB, buonacorsi2020PRB, fujita2017NPJQI, villavicencio2013PRB, zhang2013PRB, Zhao2016SR, Zhao2018SR, zajac2016PRapplied}.  With spin and spin-charge hybrid qubits having been demonstrated as hopeful candidates for foundational building blocks of future quantum processors \citep{tyryshkin2006JoPCM, bluhm2011NatPhys, petersson2012Nature, tyryshkin2012NatMat, Saeedi2013Science, Veldhorst2014NatNano, Sigillito2015PRL}, accurately characterizing tunnel coupling between quantum dots is an imperative task in characterizing these qubits.

A robust approach to detect tunnel coupling in a DQD was developed more than a decade ago \citep{DiCarlo2004PRL} based on measuring the charge distribution of the DQD in thermal equilibrium as a function of the interdot detuning, then fits a two-level (2L) model \citep{shevchenko2010PhysRep} to obtain the tunnel coupling between the two single-dot ground states.  This measurement technique is particularly successful for a GaAs DQD, where excited orbital states are generally several meV above the ground state while experimental temperature is kept at about 100 mK (for a thermal energy of $\sim 10 \mu$eV), so that the 2L model including only the single-dot ground states works perfectly \cite{DiCarlo2004PRL, petta2004PRL}.

In recent years, studies of spin qubits have focused on Si QDs because of their superior coherence properties \citep{tyryshkin2006JoPCM, bluhm2011NatPhys, petersson2012Nature, pla2012Nature, tyryshkin2012NatMat, Saeedi2013Science, Veldhorst2014NatNano, Sigillito2015PRL}.  The thermal equilibrium charge sensing technique has been widely used to measure tunnel coupling between the ground orbital states of two neighboring dots \citep{Eenink2019NanoLett, Sigillito2019PRApplied, mills2019APL, simmons2009NanoLett, yoneda2020arXiv, jones2019PRApplied, ota2010APL, stehlik2012PRB, petersson2010NanoLett}.  However, in Si-based QDs, the valley degree of freedom introduces extra energy levels a fraction of meV above the ground states, making the Si DQDs better described as a four-level (4L) system instead of a 2L system. There are also two relevant tunnel coupling parameters instead of one.  Clearly, a 2L model cannot represent all the relevant properties of a Si DQD.  It is not even clear whether it is capable of consistently producing accurate measurement of the ground state tunnel coupling.
Although alternative schemes such as spin-cavity coupling have been employed to successfully measure tunnel coupling \citep{mi2017PRL}, the thermal equilibrium charge sensing technique will still be the most easily accessible and widely used in the foreseeable future. As such, it is important to develop an updated procedure of measuring tunnel coupling in a Si DQD that accounts for the valley dynamics.

\begin{figure}
	\begin{centering}
		\includegraphics[width=1\linewidth]{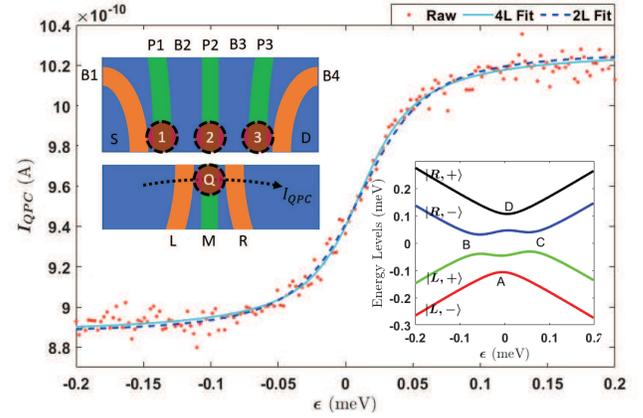}
		\par\end{centering}
	\caption{\label{fig:1}The $I_{QPC}(\epsilon)$ curve measured form experiment
		(red-doted markers) and the best fit curves. top-left inset: Schematic
		diagram of triple dot configuration. QDs ``1,2,3'' are confined
		under the plunger gates ``P1, P2, P3''. Dot ``1-2'' and ``2-3''
		forms two DQD systems. Barrier gates ``B1'' to ``B4'' control
		the tunneling strength. Dot ``Q'' plays the role of charge sensor.
		Bottom-right inset: Typical energy levels of DQD.}
\end{figure}

In this paper, we develop a four-level (4L) model for the thermal equilibrium charge sensing measurement of tunnel coupling for a Si DQD.  Specifically, we develop a numerical 4L fitting procedure for both intra- and inter-valley tunnel couplings of a Si DQD.  We also derive a perturbative 4L fitting formula, which speeds up the fitting procedure dramatically while maintaining high degree of accuracy under most conditions.  We apply the updated fitting procedures on multiple sets of data obtained in a linear Si/SiGe triple quantum dot (schematic diagram in Fig.~\ref{fig:1}) provided by Adam Mills and Jason Petta \citep{mills2019APL, mills2019NC}, and produce consistent fitting for sometimes noisy experimental data.  We compare the results from these new fitting procedures with the conventional 2L-based approach, and find significant differences under common conditions.  For example, in a particular DQD, we observe an average of 46\% difference in the intra-valley tunnel coupling between the 2L fitting and 4L fitting. For a particular set of data, the two models make totally contradictory predictions on tunnel couplings, with the 4L prediction more consistent with the experimental procedure. These examples clearly illustrate the advantages and necessity of the 4L model in obtaining a reliable estimate of tunnel coupling in a Si DQD.  Lastly, we analyze the robustness of our fitting procedure and identify possible errors.

\section*{Results}
\subsection*{Charge distribution in a four-level model}

The six-fold degeneracy of the Si conduction band is lifted by the growth-direction (nominally the $z$ direction) confinement near an interface, which leaves two of the bulk valleys with lower energy, denoted as $|z\rangle$ and $|\bar{z}\rangle$ states.  Scattering at the interface further couples $|z\rangle$ and $|\bar{z}\rangle$ states \citep{ando1982RMP, culcer2010PRB, culcer2010PRB2}, leading to the valley eigenstates $|\pm\rangle = \frac{1}{\sqrt{2}} \left( |z\rangle \pm e^{i\phi}|\bar{z} \right)$, where the phase $\phi$ is determined by the interface scattering matrix element.  The energy splitting $2|\Delta|$ between $|\pm\rangle$ states is called the valley splitting, and is typically 0.1 to 0.2 meV in a SiMOS quantum dot and $\lesssim 0.1$ meV in a Si/SiGe dot.  Compared to the few meV orbital excitation energy in these quantum dots due to in-plane confinement, valley splitting is much smaller, making it reasonable to neglect intra-valley orbital excitation but include both valleys when considering charge distribution in thermal equilibrium at low temperatures.

A minimal model for the low-energy single-electron charge distribution and dynamics of a Si DQD should thus include the ground orbital state in each dot (denoted as $|L\rangle$ and $|R\rangle$ for left and right dot), together with both valley eigenstates, leading to four basis states: $\{|L,+\rangle$, $|L,-\rangle$, $|R,+\rangle$, and $|R,-\rangle\}$.  Considering that interdot barrier is generally a smooth variation in electrical potential at a length scale much larger than the lattice constant, tunneling is only allowed between states in the same bulk valley: $\langle L,z|H|R,z\rangle=t_{C}$ is finite while $\langle L,z|H|R,\bar{z}\rangle=0$.  Using the four basis states $|D,\pm\rangle=\frac{1}{\sqrt{2}}(|D,z\rangle\pm e^{i\phi_{D}} |D,\bar{z})$ ($D=L,R$) with local phases $\phi_{D}$, the single-electron Hamiltonian in the Si DQD can be expressed as

\begin{equation}
H=\left[\begin{array}{cccc}
\epsilon+|\Delta_{L}| & 0 & t_{+} & t_{-}\\
0 & \epsilon-|\Delta_{L}| & t_{-} & t_{+}\\
t_{+}^{*} & t_{-}^{*} & -\epsilon+|\Delta_{R}| & 0\\
t_{-}^{*} & t_{+}^{*} & 0 & -\epsilon-|\Delta_{R}|
\end{array}\right] \,.\label{eq:H2}
\end{equation}
Here $\epsilon$ is the interdot detuning, $|\Delta_{L,R}|$ are the L/R valley splittings, $t_{\pm}=\frac{1}{2}t_{C}[1 \pm e^{-i\delta\phi}]$ are the intra- and inter-valley (here ``valley'' means the valley eigenstates $|\pm\rangle$) tunnel couplings, respectively, and $\delta\phi = \phi_{L}-\phi_{R}$ is the valley phase difference between the two dots.

Hamiltonian (\ref{eq:H2}) can be numerically diagonalized at any given detuning $\epsilon$ to obtain the eigenvalues $E_{i}$ and the corresponding eigenstates $|\Psi_{i}\rangle$ ($i=1,2,3,4$).  It can also be diagonalized analytically, though the general expressions are cumbersome and not transparent.  If we treat inter-valley tunneling as a perturbation, on the other hand, we can  obtain simple analytical expressions for $E_{i}$ and $|\Psi_{i}\rangle$, (see the ``Methods'' section), which can be used in a fitting procedure much more conveniently.  Specifically, the left-dot charge distribution for each eigenstate $|\Psi_{i}\rangle$ takes the form
\begin{align}
P_{L1} & =\cos^{2}\frac{\Theta_{1}}{2}\sin^{2}\frac{\theta_{-}}{2}+\sin^{2}\frac{\Theta_{1}}{2}\cos^{2}\frac{\theta_{+}}{2},\nonumber \\
P_{L2} & =\sin^{2}\frac{\Theta_{2}}{2}\cos^{2}\frac{\theta_{-}}{2}+\cos^{2}\frac{\Theta_{2}}{2}\sin^{2}\frac{\theta_{+}}{2},\nonumber \\
P_{L3} & =\cos^{2}\frac{\Theta_{2}}{2}\cos^{2}\frac{\theta_{-}}{2}+\sin^{2}\frac{\Theta_{2}}{2}\sin^{2}\frac{\theta_{+}}{2},\nonumber \\
P_{L4} & =\sin^{2}\frac{\Theta_{1}}{2}\sin^{2}\frac{\theta_{-}}{2}+\cos^{2}\frac{\Theta_{1}}{2}\cos^{2}\frac{\theta_{+}}{2}.\label{eq:PLi}
\end{align}
Here $\tan\Theta_{1} = \frac{|t_{-}|}{E_{+} + E_{-}+\Delta_{+}}$, $\tan\Theta_{2} = \frac{|t_{-}|}{E_{+} + E_{-}-\Delta_{+}}$, and $\tan\theta_{\pm} = \frac{|t_{+}|}{\epsilon \pm \Delta_{-}}$ ($\theta_{\pm}\in[0,\pi]$), with $E_{\pm} = \sqrt{(\epsilon \pm \Delta_{-})^{2}+|t_{+}|^{2}}$ and
$\Delta_{\pm} = \frac{1}{2}(|\Delta_{L}| \pm |\Delta_{R}|)$.  The expressions given in Eq.~(\ref{eq:PLi}) become exact if $|\Delta_L| = |\Delta_R|$.  A more detailed study in the ``Methods'' section shows that the approximation underlying Eq.~(\ref{eq:PLi}) is valid in most regions of the parameter space.  For example, there is only a 4\% error when $|\Delta_{L}|$ and $|\Delta_{R}|$ are different by 20\%.

When the single electron in the DQD is in thermal equilibrium, its density matrix is given by a thermal state $\rho=\sum_{i}\frac{1}{Z}e^{-\beta E_{i}}|\Psi_{i}\rangle\langle\Psi_{i}|$, where $\beta=\frac{1}{k_{B}T}$, $k_{B}$ is the Boltzmann constant, and $Z = \sum_i e^{-\beta E_{i}}$ is the partition function.  The total charge occupation in the left dot at temperature $T$ is then
\begin{equation}
P_{L}=\sum_{i}\frac{1}{Z}e^{-\beta E_{i}}P_{Li} \,. \label{eq:PL2}
\end{equation}
Here $P_L$ is a function of both tunnel couplings $t_{\pm}$, both valley splittings $|\Delta_L|$ and $|\Delta_R|$ (with their phase difference $\delta \phi$ already contained in $t_\pm$), and detuning $\epsilon$.  Given experimentally measured $P_L(\epsilon)$, we could thus obtain $t_\pm$ via data fitting.  In theory one could obtain the valley splittings from this fitting procedure as well, though our numerical studies below show that the results are not particularly sensitive to $|\Delta_{L,R}|$, making the information obtained from fitting less reliable.  Thus we generally treat valley splittings as known parameters.

There are two major reasons that cause different predictions between 2L and 4L theories. First, the different number of levels involved means that the thermal occupations are distributed differently.  The impact of this thermal occupation is typically limited since experimental temperature is usually about 100 mK and much smaller than valley splittings in the dots.  Obvious exceptions include cases when the valley splittings are very small (for example $\sim10\mu$eV, similar to the thermal energy), or when the temperature is much higher than usual. Second, and more importantly, all eigenstates $|\Psi_{i}\rangle$ in a Si DQD contain the valley excited states $|+\rangle$ due to the finite inter-valley tunnel coupling. The involvement of the excited valley states causes subtle changes to the state compositions, which then lead to differences in the charge distribution.

In the absence of inter-valley tunneling ($\delta\phi=0$), the two valley eigenstates decouple into their own subspaces, so that the charge distribution is reduced to an exact analogy to the two-level case in GaAs when we neglect the thermal occupation of the excited valley states 
\begin{equation}
	P_{L}=\frac{1}{2}\left[1-\frac{\epsilon-\Delta_-}{2E_{+}}\tanh\left(\frac{E_{+}}{2k_{B}T}\right)\right] \,.\label{eq:PL1}
\end{equation}
This is just the fitting formula in Ref.~\onlinecite{DiCarlo2004PRL} with an $\epsilon$ shift caused by asymmetric valley splittings. If we impose a further condition that the valley splittings are symmetric ($\Delta_- = 0$), the thermal occupation of the excited valley states would have the same left-right distribution as the ground valley states, so that the 4L theory we develop here would become identical to the conventional 2L model. In other words, under the condition
\begin{equation}
   \delta \phi=0 \quad \& \quad |\Delta_L|=|\Delta_R| \label{eq:reduce2L}
\end{equation}
Eq.~(\ref{eq:PLi}) and Eq.~(\ref{eq:PL2}) would lead exactly to the 2L fitting formula in Ref.~\onlinecite{DiCarlo2004PRL}, as is Eq.~(\ref{eq:PL1}).

\subsection*{\label{sec:Curve-fitting-procedure}Extracting tunnel couplings from charge distributions}

The functional forms for charge distribution in a Si DQD given by Eqs.~(\ref{eq:PLi}) and (\ref{eq:PL2}) allow us to obtain tunnel couplings $t_\pm$ (or $t_C$ and $\delta \phi$) between the two dots by fitting experimentally measured $P_L(\epsilon)$, similar to the procedure given in Ref.~\onlinecite{DiCarlo2004PRL}.

As we discussed above, the charge distribution $P_L$ is a function of multiple parameters and variables: $P_L = P_L(\Delta_L, \Delta_R, t_+, t_-, \epsilon)$.  To obtain more constrained and reliable knowledge of the tunnel couplings, the valley splittings $\Delta_L$ and $\Delta_R$ should be known beforehand, for example through the spin relaxation hot spot for each dot \cite{yang2013NatComm}.  If $|\Delta_{L,R}|$ are not known a priori, one can use an estimate instead, without creating significant errors. A detailed discussion of the consequences of not knowing these splittings is given in the ``Methods'' section.

Experimentally, what is measured is the charge sensor (for example a quantum point contact, or QPC) current as a function of the interdot detuning: $I_{QPC}(\epsilon)$.  The current is usually assumed to be linearly related to the charge distribution in the DQD \cite{DiCarlo2004PRL}:
\begin{equation}
I_{QPC}=I_{0}+\delta I\cdot P_{L}(\epsilon)+\delta I_{noise} \,, \label{eq:IQPC}
\end{equation}
where $I_{0}$ is the background current (setting a reference point), $\delta I$ is the linear conversion ratio, and $\delta I_{noise}$ is the noise in the $I_{QPC}$ measurement. The first two parameters are part of the fitting procedure and the impact of $\delta I_{noise}$ will be discussed in the ``Methods'' section.  In addition, the inter-dot detuning may also have a background voltage, i.e., $\epsilon_{m}=\epsilon_{0}+\epsilon$, where $\epsilon_{m}$ is the value measured in the experiment and $\epsilon_{0}$ is a reference shift.

Our fitting procedure thus consists of the following steps.  First we use Eq.~(\ref{eq:PL2}) or Eq.~(\ref{eq:PL1}) (for 4L or 2L fitting respectively) to generate a theoretical curve $I_{th} (\epsilon)$ with a set of candidate fitting parameters such as $t_{\pm}$.  We then calculate the deviation from the experimental data, and minimize it by varying the fitting parameters.  While the three parameters $I_{0}$, $\delta I$, and $\epsilon_{0}$ are part of the fitting parameter set, they take up different roles compared to $t_\pm$. The tunnel couplings $t_{\pm}$ determine the ``shape'' of the curve, while these three parameters determine the positions of the curve. In particular, $\epsilon_{0}$ determines the shift in the horizontal (detuning) direction, $I_{0}$ determines the vertical shift, while $\delta I$ is a scaling factor. None of them contributes to the shape or curvature of the curve near $\epsilon=0$, which is determined by $t_{\pm}$. Therefore, they can be obtained separately from the main fitting parameters $t_{\pm}$. One can follow an adaptive fitting procedure that fit these two groups of parameters in turn until they converge to steady values, respectively. A discussion about the fitting inaccuracy caused by errors in $I_{0},\delta I$, and $\epsilon_{0}$ can be found in the ``Methods'' section, particularly in Fig.~\ref{fig:guessI0dIep}.

\subsection*{\label{sec:FitExp}Fitting actual experimental data: an example}

With the procedure described above, we examine some experimental data acquired during the tune-up of a linear array of 9 QDs used to demonstrate charge shuttling \citep{mills2019APL, zajac2016PRapplied, mills2019NC}. The measurements were performed on a triple dot schematically shown in Fig.~\ref{fig:1}. It is part of a Si/SiGe 9-dot array with three QPCs as charge sensors \citep{mills2019APL, zajac2016PRapplied, mills2019NC}.  The experimental temperature is at $T=50$ mK \cite{mills2019APL} and the valley splittings $|\Delta_{L}|$ and $|\Delta_{R}|$ are estimated to be around 66-74 $\mu$eV from spin measurements in the same device \cite{Mills2020PrivateCommu}. For each DQD, the $I_{QPC}(\epsilon)$ curve are measured with four different barrier gate voltage $V_{B2}$ (or $V_{B3}$). One set of data, together with our fitting curve, is shown in Fig.~\ref{fig:1}.  All other data sets and fitting curves are shown in Fig.~\ref{fig:BestFitAll} in the ``Methods'' section.

% Alternative presentations: combine tab1 and tab2.

\begin{table}
	\begin{tabular}{|c|c|c|c|c|}
		\hline 
		QD 1-2  & (a)  & (b)  & (c)  & (d)\tabularnewline
		\hline 
		\hline 
		2L $|t_{+}|$  & $24\pm0.6$  & $43\pm1.1$  & $53\pm1.0$  & $70\pm2.5$\tabularnewline
		\hline 
		4L $|t_{+}|$ (N.) & $20\pm1.2$  & $32\pm1.9$  & $37\pm2.6$  & $37\pm6.0$\tabularnewline
		\hline 
		4L $|t_{+}|$ (F.) & $20\pm1.1$  & $33\pm1.8$  & $37\pm2.6$  & $38\pm5.7$\tabularnewline
		\hline 
		4L $|t_{-}|$  & $39\pm4.7$  & $64\pm4.8$  & $76\pm5.9$  & $112\pm12$\tabularnewline
		\hline 
		4L $\delta\phi$ (rad) & $2.2\pm0.05$  & $2.2\pm0.03$  & $2.2\pm0.04$  & $2.5\pm0.06$\tabularnewline
		\hline 
		\hline 
		QD 2-3 & (e) & (f) & (g) & (h)\tabularnewline
		\hline 
		\hline 
		2L $|t_{+}|$ & $22\pm0.6$ & $41\pm0.8$ & $44\pm0.8$ & \textbf{$\bm{36}\pm1.3$}\tabularnewline
		\hline 
		4L $|t_{+}|$ (N.) & $22\pm0.7$ & $41\pm1.7$ & $44\pm1.6$ & $26\pm2.5$\tabularnewline
		\hline 
		4L $|t_{+}|$ (F.) & $22\pm0.8$ & $41\pm1.6$ & $44\pm1.6$ & $26\pm2.5$\tabularnewline
		\hline 
		4L $|t_{-}|$ & $0\pm9.6$ & $0\pm14$ & $0\pm14$ & $62\pm7.1$\tabularnewline
		\hline 
		4L $\delta\phi$ (rad) & $0\pm0.22$ & $0\pm0.20$ & $0\pm0.18$ & $2.3\pm0.04$\tabularnewline
		\hline 
	\end{tabular}
	
	\caption{\label{tab:1}
		Best fitting parameters for tunnel couplings. Data set
		(a)-(d) are measured from dot 1-2 and fitted with $|\Delta_{L}|=66$
		$\mu$eV, $|\Delta_{R}|=74$ $\mu$eV. Data set (e)-(h) are measured
		from dot 2-3 and fitted with $|\Delta_{L}|=74$ $\mu$eV, $|\Delta_{R}|=74$
		$\mu$eV. $|t_{\pm}|$ units are $\mu$eV. (N.) means using numerical
		diagonalization, (F.) means using Eq. (\ref{eq:PLi}).}
\end{table}

The tunnel couplings and other parameters we obtained from data fitting are summarized in Table~\ref{tab:1}.  In particular, the inter-dot valley phase difference $\delta\phi$  for QD 1-2  in Table~\ref{tab:1} is roughly a constant under different applied $V_{B2}$, which implies that varying $V_{B2}$ only changes the interdot barrier height, but does not cause the dots to shift to any significant degree.  Consequently, in the 4L model only $t_C$ depends on $V_{B2}$, while $\delta \phi$ does not.

The tunnel couplings are obtained with different fitting formulas: (A). Eq.~(\ref{eq:PL1}), labeled as ``2L'', (B). Eq.~(\ref{eq:PL2}) and (\ref{eq:PLi}), labeled as ``4L (F.)'', (C). Eq.~(\ref{eq:PL2}), with a numerical diagonalization
of $H$ to calculate $P_{Li}$, labeled as ``4L (N.)''. According to the fitting results, the analytical formula Eq.~(\ref{eq:PLi})
provides almost the same results as numerically diagonalizing the Hamiltonian. The average difference between 4L $|t_{+}|$ (F.) and 4L $|t_{+}|$ (N.) is only \textbf{0.9\%}, which indicates that Eq.~(\ref{eq:PLi}) is very accurate here. Additional discussion about the accuracy of Eq.~(\ref{eq:PLi}) will be presented in following subsections as well as in the ``Methods'' section.

In comparison, the 2L fitting results are quite different from the 4L results: the average difference between 2L $|t_{+}|$ and 4L $|t_{+}|$ for QD 1-2 is \textbf{46\%} across the different barrier heights in Table~\ref{tab:1}, with 2L model consistently producing larger tunnel splittings.  Qualitatively, this deviation is due to the fact that in the 2L model we are using a single excited level to represent the effect of three excited levels in the 4L model.  As such this single excited state needs to be above the first excited state but lower than the third excited state in the 4L model, so that $|t_+|$ in the 2L model has to be larger than that in the 4L model. The error bars are obtained by numerically generating various stochastic realization of $\delta I_{noise}$ with the same standard deviation as the measured data and then fitting all realizations. Here we assume the uncertainty mainly comes from the noise in the $I_{QPC}$ signal as shown in Fig.~\ref{fig:1}, which is the conclusion from Ref.~\citep{DiCarlo2004PRL} as well.

We perform the curve fitting with several combinations of $|\Delta_{L}|$ and $|\Delta_{R}|$ within the estimated range of 66 to 74 $\mu$eV. The parameters presented in Table~\ref{tab:1} are chosen because they produce the most consistent fitting results for $\delta\phi$. The fitting results for other $|\Delta_{L,R}|$ are shown in Fig.~\ref{fig:fittingvsdphi} in the ``Methods'' section.

The best fittings for QD 2-3 are shown in Table~\ref{tab:1}, data set (e)-(h).  A notable contradiction arises for (h).  The 2L theory predicts $t_C=|t_{+}| = 36 \mu$eV for (h), which is smaller than $t_C=|t_{+}|=44 \mu$eV for (g), even though the increase in $V_{B3}$ from (g) to (h) should cause the barrier height to decrease and tunnel coupling to increase \citep{Mills2020PrivateCommu}.  This abnormality does not show up in the 4L theory, which suggests that $t_{C}=\sqrt{|t_+|^2+|t_-|^2}=67 \mu$eV  for (h), larger than $t_{C} = 44 \mu$eV for (g).  However, the 4L result of $\delta \phi$ for (h) is remarkably different from other fitting results for QD 2-3, as if the dots have shifted so that at least one of the dots has a significantly different valley phase.  Indeed multiple factors could cause this change in $\delta \phi$.  A physical reason such as possible interface steps in the DQD \citep{tariq2019PRB} could lead to such a shift.  Other reasons, such as the relatively noisy data (see Fig.~\ref{fig:BestFitAll} in the ``Methods'' section) or non-linear effect in Eq.~(\ref{eq:IQPC}) \citep{DiCarlo2004PRL}, could cause a change in $\delta \phi$, too. Under imperfect conditions, such as a large $\delta I_{noise}$ for (h), both 2L and 4L theory may fail to provide accurate fitting results, although the 2L prediction is qualitatively worse since it contradicts the experimental procedure.

The example given in Fig.~\ref{fig:1} and Table~\ref{tab:1} clearly shows that our 4L model is a much better representation of a Si DQD, and provides more reliable knowledge of the tunnel coupling in the DQD compared to the 2L model that has been widely used so far.  Qualitatively, the 2L theory only includes the intra-valley tunneling reflected by anti-crossing ``A'' in Fig.~\ref{fig:1}, resulting in a simple form of ground state charge distribution $P_L=\sin ^2 \frac{\theta_-}{2}$. The inter-valley tunneling reflected by anti-crossings ``B'' and ``C'' are taken into account in our 4L theory.  They provide a correction to the 2L theory as represented by the factors $\cos^{2}\frac{\Theta_{1}} {2}$ and $\sin^{2}\frac{\Theta_{1}} {2}$ ($t_{-}$ dependent) in Eq.~(\ref{eq:PLi}). The details are presented in the ``Methods'' section and the magnitude of the 4L corrections from ``B'' and ``C'' will be discussed further in the next subsection.

\subsection*{\label{sec:Accuracy}Comparison between two-level and four-level models}

Our results above show that 2L fitting returns quite different numbers from the 4L fitting.  Nevertheless, it is still valuable to investigate the differences between the two models in a wide range of experimental conditions (under various $|t_{\pm}|$, $|\Delta_{L,R}|$, $T$, etc.), to better clarify their applicability in experimental studies of Si DQDs, which is the subject of this subsection. 

In order to investigate a certain set of parameters ($|t_{\pm}|$, $|\Delta_{L,R}|$, $T$) other than the measured ones shown in Table~\ref{tab:1}, we employ the Hamiltonian (\ref{eq:H2}) to calculate a ``pseudo-curve'' $I_{QPC}(\epsilon)$ theoretically with given parameters. It represents an expected $I_{QPC}(\epsilon)$ curve measured in experiments with particular parameters. Then, we apply the procedure we proposed to fit this ``pseudo-curve'' and compare the fitted parameters with the original parameters used to generate the curve.

\begin{figure}
	\includegraphics[width=1\linewidth]{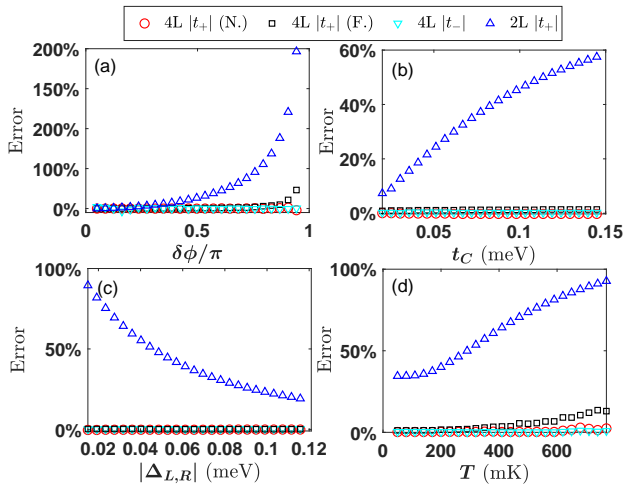}
	
	\caption{\label{fig:2}Errors for different fitting methods. The parameters
		to generate the ``pseudo-curve'' are $|\Delta_{L}|=0.066$ meV,
		$|\Delta_{R}|=0.074$ meV. $t_{C}=0.071$ meV, $\delta\phi=0.7\pi$,
		and $T=50$ mK unless specified explicitly in the figures.  As shown in the legend, the blue-upper-triangles are results from the 2L formula~(\ref{eq:PL1}), the red circles are obtained via numerical diagonalization of the 4L Hamiltonian, while the black squares are from the 4L analytical formula of Eq.~(\ref{eq:PLi}).}
\end{figure}

We first discuss qualitatively the necessary condition for a 4L model description of a Si DQD.  Consider two well separated dots, when $t_{\pm}$ can be treated as perturbations in Eq.~(\ref{eq:H2}).  The eigen-states $|\Psi_{i}\rangle$ are mainly the four unperturbed states $|L,\pm\rangle$, $|R,\pm\rangle$ except near the anti-crossings. For example, when $\epsilon$ is in
the region $\epsilon_{A}\pm\Delta\epsilon$, where $\epsilon_{A}=(|\Delta_{L}|-|\Delta_{R}|)/2$ is the center of anti-crossing ``A'', the ground state would be a mixture of $|L,-\rangle$ and $|R,-\rangle$. The range of this ``mixing region'' is roughly in the order of intra-valley tunneling, $\Delta\epsilon\sim|t_{+}|$. Outside the ``mixing region'', eigen-states $|\Psi_{i}\rangle$ are not affected by the this anti-crossing and remain the unperturbed states. Similarly, anti-crossings ``B'' and ``C'' also have their own ``mixing regions'', but mix different states $|L,+\rangle \rightarrow |R,-\rangle$ or $|L,-\rangle \rightarrow |R,+\rangle$, with a different magnitude of $\Delta \epsilon \sim|t_{-}|$.  The charge distribution of the ground state is mainly determined by anti-crossing ``A'', as illustrated in Fig. \ref{fig:1}.  The impacts of ``B'' and ``C'' can be regarded as corrections. Therefore, when ``B'' and ``C'' are far away from ``A'', the impacts are limited and the dynamics is roughly a 2L dynamics with a single anti-crossing ``A''.  In other words, under the condition 
\begin{equation}
	|\epsilon_{A}-\epsilon_{B}|=|\Delta_{L}|\gg|t_{-}|\quad \& \quad|\epsilon_{A}-\epsilon_{C}|=|\Delta_{R}|\gg|t_{-}|,\label{eq:Cond}
\end{equation}
the 2L theory proposed in Ref.~\onlinecite{DiCarlo2004PRL} is still a good approximation to describe the ground state charge distribution of a Si DQD. When the condition~(\ref{eq:Cond}) is not fulfilled, on the other hand, a 4L model is necessary.

In Fig.~\ref{fig:2} we provide numerical evidences for the condition~(\ref{eq:Cond}), in the form of fitting errors' dependences on various system parameters.  Specifically, Fig.~\ref{fig:2}(a) shows the effects of $\delta\phi$ with a fixed $t_C$ and valley splittings. Clearly, Eq.~(\ref{eq:Cond}) is fulfilled when $|t_{-}|\approx0$, which requires $\delta\phi\approx0$. In this case, the valley states $|+\rangle$ does not couple to $|-\rangle$, and the 4L system is approximately reduced to a pair of 2L system.  When $\delta\phi$ is finite, on the other hand, the 2L fitting generally results in significant errors, especially when $\delta\phi\rightarrow\pi$. At this limit $|t_+| \rightarrow 0$, making the 2L model unstable and sensitive to any data noise.

Figure \ref{fig:2}~(b), (c), and (d) show the effects of tunnel coupling $t_{C}$, valley splittings $|\Delta_{L}|$ and $|\Delta_{R}|$, and temperature.  The results are all consistent with condition Eq.~(\ref{eq:Cond}).  Here $t_C$ represents tunnel coupling between bulk valleys, and is not directly measurable.  However, larger $t_{C}$ leads to larger $|t_{-}|$ for a given $\delta \phi$, making the 2L theory less reliable as the condition Eq.~(\ref{eq:Cond}) is weakened.
Similarly, when $|\Delta_{L,R}|$ is large, the fitting error by 2-level theory is significantly suppressed, while a smaller $|\Delta_{L,R}|$ leads to overlapping of mixing regions for anti-crossings A, B, and C, and the 2-level theory fails. 
The large errors of the 2L fitting
at higher temperatures are also expected, as there is only one excited state with a simple $\epsilon$ dependence as opposed to three excited states with much more complex $\epsilon$ dependence.

One advantage of the 4L theory is that it automatically extracts inter-valley tunnel coupling $t_-$ from the ground state charge distribution, as shown with the cyan triangles in all four subplots of Fig.~\ref{fig:2}.  We also note that
the approximate charge distributions given by the analytical expressions in Eq.~(\ref{eq:PLi}) is very
accurate except in a very small region when $\delta\phi\approx\pi$ in
Fig.~\ref{fig:2}~(a). Therefore, Eq.~(\ref{eq:PLi}) are a perfect approximation in most cases, allowing a much faster fitting calculation compared to a fully numerical procedure. A more detailed study on the accuracy
of Eq.~(\ref{eq:PLi}) is in the ``Methods'' section and Fig.~\ref{fig:AccFormula}.

There are also other practical factors affecting the accuracy of the fitting such as the inaccurate fitting of the parameters $\delta I$, $I_0$, $\epsilon_0$, insufficient information of $|\Delta_{L,R}|$, and signal noise in $I_{QPC}$. The impacts of all these factors are discussed in the ``Methods'' section.

\section*{Discussion}

In this paper, we present a four-level model that can extract tunnel coupling information in a Si double quantum dot via measurement of charge distribution of the double dot in thermal equilibrium.  In essence, we have adapted the protocol originally proposed and used for GaAs double dot \cite{DiCarlo2004PRL} to a Si DQD by including the valley-orbit coupling and dynamics.  We demonstrate the efficiency and robustness of our model and the associated fitting procedure by applying it to experimental data collected in a pair of Si DQD $I_{QPC}(\epsilon)$ \citep{mills2019APL}.  The results clearly demonstrate the superiority of the 4L model compared to the conventional two-level model used in the original proposal, with the 2L model produces almost 50\% larger tunnel coupling $t_+$, not to mention that only by the 4L model can one extract any information on the inter-valley tunnel coupling $t_-$.  In addition to directly diagonalizing the 4L model Hamiltonian in the fitting procedure, we have also derived a set of approximate formula with the assumption that $t_-$ can be treated perturbatively.  Our numerical results show that the approximate formula perform nearly perfectly in the vast majority of parameter regimes, with the only exception near the point where the inter-dot valley phase difference is $\pi$. 
Lastly, we compare the performance of the 2L and 4L models, and clarify the condition under which 4L model is necessary. 
In short, our 4L model for a Si DQD provides much better accuracy in extracting the intravalley tunnel coupling $t_+$ from a charge distribution measurement, while carrying the extra benefit of also extracting inter-valley tunnel coupling $t_-$. We hope that the proposed protocol can help experimentalists to measure tunnel couplings for a Si DQD more accurately and efficiently.

\section*{Methods}

\subsection*{Theoretical charg\label{sec:App1}e distribution}

In our approximate treatment, we consider inter-valley tunnel coupling as a perturbation, while include intra-valley tunnel couplings in the unperturbed Hamiltonian. In essence we take a DQD with a completely smooth interface as our starting point.  The Hamiltonian
(\ref{eq:H2}) can thus be split into two parts 
\begin{equation}
	H=H_{0}+H_{1},
\end{equation}
where
\begin{equation}
	H_{0}=\left[\begin{array}{cccc}
		\epsilon+|\Delta_{L}| & 0 & t_{+} & 0\\
		0 & \epsilon-|\Delta_{L}| & 0 & t_{+}\\
		t_{+}^{*} & 0 & -\epsilon+|\Delta_{R}| & 0\\
		0 & t_{+}^{*} & 0 & -\epsilon-|\Delta_{R}|
	\end{array}\right],
\end{equation}
\begin{equation}
	H_{1}=\left[\begin{array}{cccc}
		0 & 0 & 0 & t_{-}\\
		0 & 0 & t_{-} & 0\\
		0 & t_{-}^{*} & 0 & 0\\
		t_{-}^{*} & 0 & 0 & 0
	\end{array}\right].
\end{equation}
The eigen-energies and eigen-states of $H_{0}$ are
\begin{equation}
	E_{1,\pm}=\pm\Delta_{+}-E_{\pm}
\end{equation}
\begin{equation}
	E_{2,\pm}=\pm\Delta_{+}+E_{\pm}
\end{equation}
where $E_{\pm}=\sqrt{(\epsilon\pm\Delta_{-})^{2}+|t_{+}|^{2}}$ and
$\Delta_{\pm}=\frac{1}{2}(|\Delta_{L}|\pm|\Delta_{R}|)$, and the corresponding
eigen vectors are 
\begin{equation}
	|\psi_{1,\mp}\rangle=\cos\frac{\theta_{\mp}}{2}|R,\mp\rangle-e^{-i\delta\phi/2}\sin\frac{\theta_{\mp}}{2}|L,\mp\rangle
\end{equation}
\begin{equation}
	|\psi_{2,\mp}\rangle=e^{i\delta\phi/2}\sin\frac{\theta_{\mp}}{2}|R,\mp\rangle+\cos\frac{\theta_{\mp}}{2}|L,\mp\rangle
\end{equation}
where $\tan\theta_{\mp}=\frac{|t_{+}|}{\epsilon\mp\Delta_{-}}$ ($\theta_{\mp}\in[0,\pi]$). 

When the inter-valley tunneling $|t_{-}|$ is finite, the Hamiltonian
$H$ can be rewritten in the new basis $\{|\psi_{1,\mp}\rangle,|\psi_{2,\mp}\rangle\}$.
The matrix representation of $H_{0}$ becomes diagonal and the matrix
elements of $H_{1}$ can be obtained as, for example, 
\begin{equation}
	\langle\psi_{1,-}|H_{1}|\psi_{1,+}\rangle=-i|t_{-}|\sin\left(\frac{\theta_{-}}{2}-\frac{\theta_{+}}{2}\right)\label{eq:sin}
\end{equation}
\begin{equation}
	\langle\psi_{1,-}|H_{1}|\psi_{2,+}\rangle=t_{-}^{*}\cos\left(\frac{\theta_{-}}{2}-\frac{\theta_{+}}{2}\right)
\end{equation}

When $|\Delta_{L}|=|\Delta_{R}|$, $\theta_{-}=\theta_{+}$.  As a result,
$\cos\left(\frac{\theta_{-}}{2}-\frac{\theta_{+}}{2}\right) = 1$ and
$\sin\left(\frac{\theta_{-}}{2}-\frac{\theta_{+}}{2}\right) = 0$.
In the new basis $\{|\psi_{1,\mp}\rangle,|\psi_{2,\mp}\rangle\}$,
the rotated full Hamiltonian $\tilde{H}$ can be written as 
\begin{equation}
	\tilde{H}=\left[\begin{array}{cccc}
		-\Delta_{+}-E_{-} & 0 & 0 & t_{-}^{*}\\
		0 & -\Delta_{+}+E_{-} & t_{-}^{*} & 0\\
		0 & t_{-} & \Delta_{+}-E_{+} & 0\\
		t_{-} & 0 & 0 & \Delta_{+}+E_{+}
	\end{array}\right]
\end{equation}
The eigen-energies are then
\begin{equation}
	E_{1}=\frac{1}{2}\left(E_{+}-E_{-}-\sqrt{(E_{+}+E_{-}+2|\Delta_{+}|)^{2}+|t_{-}|^{2}}\right)
\end{equation}

\begin{equation}
	E_{2}=\frac{1}{2}\left(E_{-}-E_{+}-\sqrt{(E_{+}+E_{-}-2|\Delta_{+}|)^{2}+|t_{-}|^{2}}\right)
\end{equation}

\begin{equation}
	E_{3}=\frac{1}{2}\left(E_{-}-E_{+}+\sqrt{(E_{+}+E_{-}-2|\Delta_{+}|)^{2}+|t_{-}|^{2}}\right)
\end{equation}

\begin{equation}
	E_{4}=\frac{1}{2}\left(E_{+}-E_{-}+\sqrt{(E_{+}+E_{-}+2|\Delta_{+}|)^{2}+|t_{-}|^{2}}\right)
\end{equation}
and the corresponding eigen-states are
\begin{equation}
	|\Psi_{1}\rangle=e^{i\phi}\cos\frac{\Theta_{1}}{2}|\psi_{1,-}\rangle-\sin\frac{\Theta_{1}}{2}|\psi_{2,+}\rangle\label{eq:psi1}
\end{equation}

\begin{equation}
	|\Psi_{2}\rangle=-e^{i\phi}\sin\frac{\Theta_{2}}{2}|\psi_{2,-}\rangle+\cos\frac{\Theta_{2}}{2}|\psi_{1,+}\rangle
\end{equation}

\begin{equation}
	|\Psi_{3}\rangle=e^{i\phi}\cos\frac{\Theta_{2}}{2}|\psi_{2,-}\rangle+\sin\frac{\Theta_{2}}{2}|\psi_{1,+}\rangle
\end{equation}

\begin{equation}
	|\Psi_{4}\rangle=e^{i\phi}\sin\frac{\Theta_{1}}{2}|\psi_{1,-}\rangle+\cos\frac{\Theta_{1}}{2}|\psi_{2,+}\rangle\label{eq:psi4}
\end{equation}
where $\tan\Theta_{1}=\frac{|t_{-}|}{E_{+}+E_{-}+\Delta_{+}}$ and
$\tan\Theta_{2}=\frac{|t_{-}|}{E_{+}+E_{-}-\Delta_{+}}$. 
The left-dot charge distribution $|\langle L,-|\Psi_{i}\rangle|^{2}+|\langle L,+|\Psi_{i}\rangle|^{2}$
for these four eigen-states are in the form of Eq.~(\ref{eq:PLi}).

\subsection*{Accuracy of the approximate solution\label{sec:App2}}

\begin{figure}
	\includegraphics[width=1\linewidth]{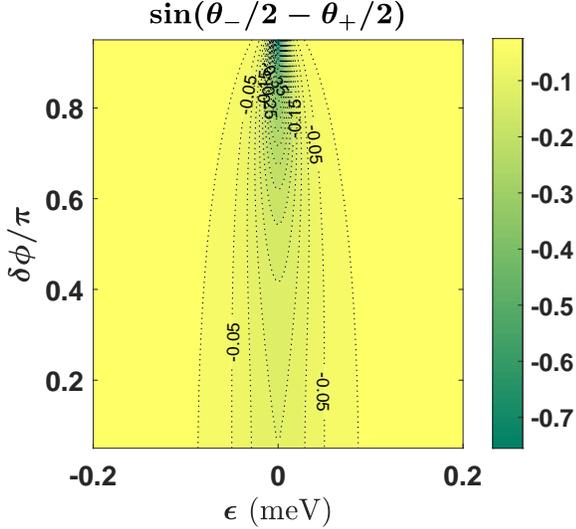}
	
	\caption{\label{fig:AccFormula}Accuracy of approximate diagonalization under
		different conditions. The valley splittings are $|\Delta_{L}|=0.045$
		meV, $|\Delta_{R}|=0.055$ meV.}
\end{figure}

The eigen-states (\ref{eq:psi1}-\ref{eq:psi4})
are obtained when $|\Delta_{L}|=|\Delta_{R}|$. Practically, $|\Delta_{L}|$ is usually not identical to $|\Delta_{R}|$.
However the charge distributions in Eq.~(\ref{eq:PLi}) remain a
good approximation. This is because in most cases the nearby dots
have similar valley splittings $|\Delta_{L}|\approx|\Delta_{R}|$,
which makes $\sin\left(\frac{\theta_{-}}{2}-\frac{\theta_{+}}{2}\right)\approx0$.
As such, the term $\langle\psi_{1,-}|H_{1}|\psi_{1,+}\rangle$ that
we neglected is generally a small correction compared to $\langle\psi_{1,-}|H_{1}|\psi_{2,+}\rangle$.
Even if $|\Delta_{L}|$ and $|\Delta_{R}|$ is quite different, our
numerical results in Fig.~\ref{fig:AccFormula} suggest only a small
error in Eq.~(\ref{eq:PLi}).

In Fig.~\ref{fig:AccFormula}, we plot the factor $\sin\left(\frac{\theta_{-}}{2}-\frac{\theta_{+}}{2}\right)$
under different QD parameters. It shows that in most region, the factor
we have neglected in Eq. (\ref{eq:sin}) is quite small. Only in the
very special case when the phase difference $\delta\phi\rightarrow\pi$
and the detuning $\epsilon\rightarrow0$, the factor $\sin\left(\frac{\theta_{-}}{2}-\frac{\theta_{+}}{2}\right)$
is notable. Otherwise, Eq.~(\ref{eq:PLi}) is always a good approximation.
Besides, for any $\delta\phi$ and $|\Delta_{L,R}|$, the notable
deviation always appears near $\epsilon=0$. If we consider the average
of $\sin\left(\frac{\theta_{-}}{2}-\frac{\theta_{+}}{2}\right)$ over
all $\epsilon$ (because the fitting depends on the charge distribution
over all $\epsilon$, not just at $\epsilon = 0$), the average deviation is always small. For
example, when there is a 20\% difference between the two valley splittings,
$|\Delta_{-}|/|\Delta_{+}|=0.2$, the overall magnitude (average value
over different $\epsilon$) of the factor $\sin\left(\frac{\theta_{-}}{2}-\frac{\theta_{+}}{2}\right)$
is only 4\% of the factor $\cos\left(\frac{\theta_{-}}{2}-\frac{\theta_{+}}{2}\right)$.
Namely, the terms we dropped are indeed negligible even when there
is a notable difference between the two valley splittings.

\subsection*{\label{sec:App4}Details of data fitting}

\begin{figure*}[t]
	\begin{centering}
		\includegraphics[width=1\linewidth]{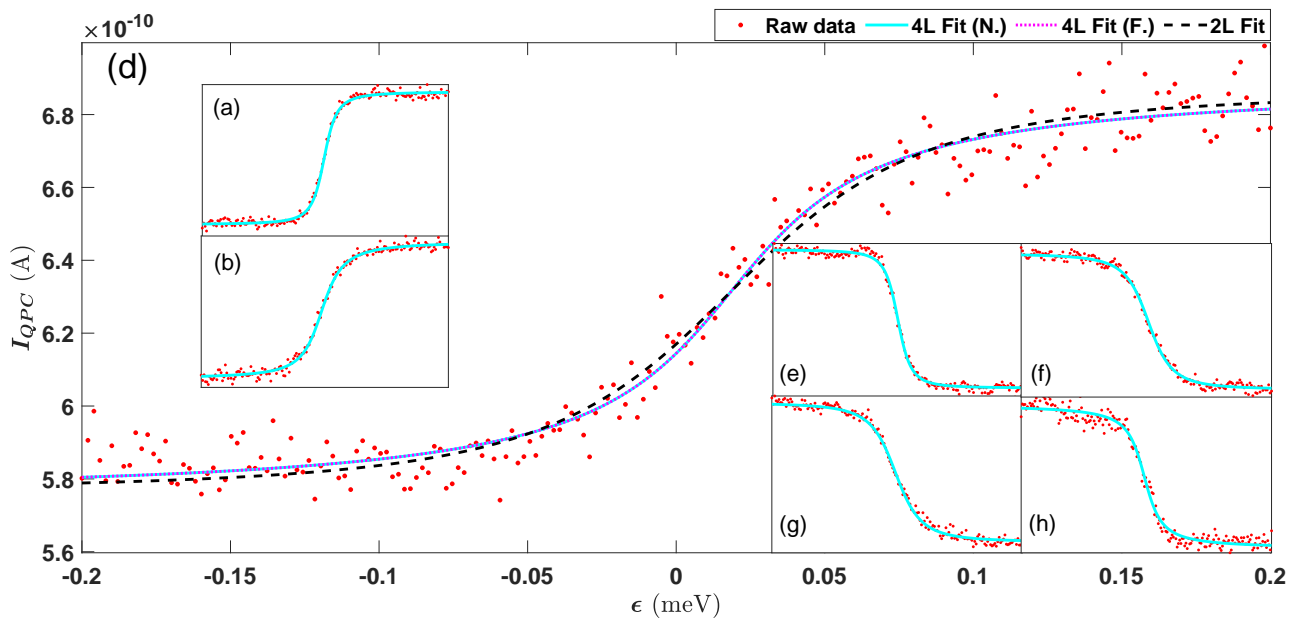} 
		\par\end{centering}
	\caption{\label{fig:BestFitAll}Curve fitting for actual data measured from
		experiment. (a), (b) and main plot (d) are obtained from the left
		two dots (dot 1 and 2) with different barrier gate voltage. (e)-(h)
		in the right-bottom corner are measured from right two dots (2 and
		3). Data set (c) is absent since it is presented 
		in Fig. \ref{fig:1}.}
\end{figure*}

\begin{figure}
	\begin{centering}
		\includegraphics[width=1\linewidth]{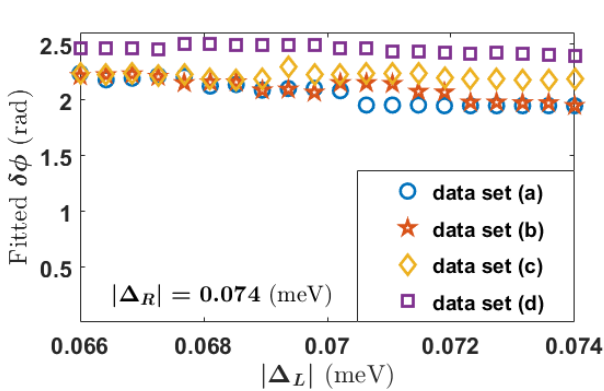} 
		\par\end{centering}
	\caption{\label{fig:fittingvsdphi}Fitting results of $\delta\phi$ by using
		various combination of $|\Delta_{L}|$ and $|\Delta_{R}|$.}
\end{figure}

In the ``Results'' section, we show all the fitting results in Table~\ref{tab:1} and the raw data of data set (c) in Fig.~\ref{fig:1}.  Here, we show all the
other 7 sets of raw data and the best fitting curves. The raw data are
extracted from the readout of the QPC sensor directly and the best
fitting curves are shown in Fig.~\ref{fig:BestFitAll}. For panels
(a), (b) and (d), the data are measured from QD~1-2, with
different barrier gate voltage $V_{B2}$ (which tunes $t_{C}$). The valley
splitting $|\Delta_{L}|$ and $|\Delta_{R}|$ are not actually measured
directly in the experiment, and are estimated to be
around 66-74 $\mu$eV \cite{Mills2020PrivateCommu}. We perform
the curve fitting with several groups of $|\Delta_{L}|$ and $|\Delta_{R}|$
ranging from 66-74 $\mu$eV, as shown in Fig.~\ref{fig:fittingvsdphi}. The data presented in Tables \ref{tab:1} is picked because the fitting results of $\delta\phi$
are relatively consistent. We choose this criterion because the only tuned parameter in the experiment
is $t_{C}$, which would generally not affect $\delta\phi$ when it is not varied too significantly. Interestingly,  Fig.~\ref{fig:fittingvsdphi} shows
that other estimates of $|\Delta_{L}|$ lead to very similar results
on $\delta\phi$. The fitting results of $\delta\phi$ is always around
2.2 rad. The value of $\delta\phi$ fitted from data set (d) has a
small difference from the results from other data sets (a), (b) and
(c). We believe this is mostly because data set (d) has larger noise
in the raw data of $I_{QPC}$, which is apparent in Fig.~\ref{fig:BestFitAll}.  In short, our fitting procedure does not seem to be overly sensitive to the choices of the valley splittings, as long as they are not too different across the two dots.

We highlight data set (d) in Fig.~\ref{fig:BestFitAll} because it
has the largest $t_{C}$, making the difference between 2L fitting
curve and 4L fitting curve clearly observable with bare eyes.
One can easily see the 4L curve fits better
to the raw data. The curve obtained by using Eq.~(\ref{eq:PLi})
almost coincide with the curve obtained by fully numerical diagonalization, illustrating the robustness of our approximate expressions.
Besides, we would also like to emphasize that the actual 2L and 4L fitting
results for (d) are quite different (almost 100\% according the results
in Table \ref{tab:1}), much larger than it seems from the two curves.

Similarly, the best curve fittings of the data measured from QD~2-3
are shown in Fig.~\ref{fig:BestFitAll}~(e)-(h).  The most interesting
result is the last panel (h). It is discussed that
the 2L theory predicts $t_{C}$(h) to be smaller than $t_{C}$(g), while
the 4L theory suggests that $t_{C}$(h) is larger than $t_{C}$(g).
Experimentally, it is expected that the true value of $t_{C}$ in
panel (h) should be larger because the gate voltage $V_{B3}$ used
to tune $t_{C}$ between dot 2-3 is increased from (g) to (h) when
the experiment is performed. Here, Fig.~\ref{fig:BestFitAll}~(h)
shows this set of data is obviously measured with a notably larger
noise than other sets. This large noise leads to more
significant error for the fitting results. However, we also
see that even with such a large noise the 4L fitting still make a
prediction which does not contradict with the experimental setup.

\subsection*{\label{sec:App0}Difference between 2L and 4L fitting curves}

\begin{figure} [h]
	\includegraphics[width=1\linewidth]{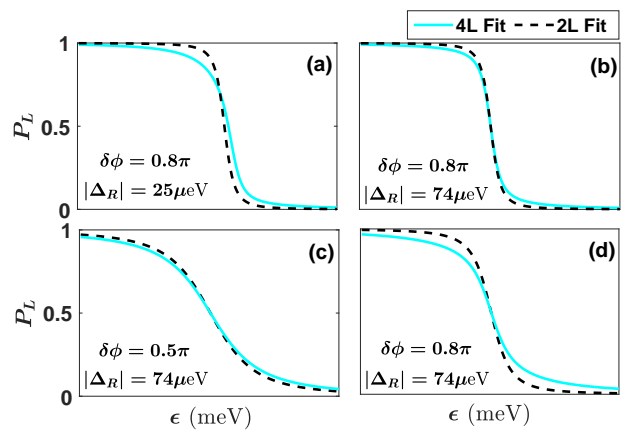}
	
	\caption{\label{fig:2Lv4L}Intuitive difference of two fitting formulas. Parameters
		are $t_{C}=50\mu$eV, $|\Delta_{L}|=74\mu$eV. $|\Delta_{R}|$ and $\delta\phi$ are given in each subplot. $\epsilon$ is ranging from $-0.2$ meV to $0.2$ meV.}
\end{figure}

The difference between the 2L fitting curve and the 4L fitting curve
in Fig.~\ref{fig:1} looks insignificant. This is mostly because the particular parameters in the experiment
happen to produce similar 2L and 4L curves, even though the corresponding tunnel coupling strengths are quite different. In Fig.~\ref{fig:2Lv4L}, we show that the differences between a 2L and a 4L fitting
curves can both be minimal and be dramatic.  For example, very different valley splittings $|\Delta_{L,R}|$ in the two dots makes the 4L curve lose its symmetry around zero detuning, while a 2L curve is always symmetric.  The
two fitting curve can also be easily distinguished when the condition (\ref{eq:Cond})
is not fulfilled. Even when two curves look very similar, they may yield very
different fitting results.  For instance, the curves in Fig.~\ref{fig:BestFitAll}~(d)
look quite similar, but there is nearly 90\% difference between the 2L and
4L fitting results on $|t_{+}|$ in Table~\ref{tab:1}.

\subsection*{\label{sec:App3}Estimation of fitting parameters $I_{0}$, $\delta I$,
	and $\epsilon_{0}$}

\begin{figure}
	\begin{centering}
		\includegraphics[width=1\linewidth]{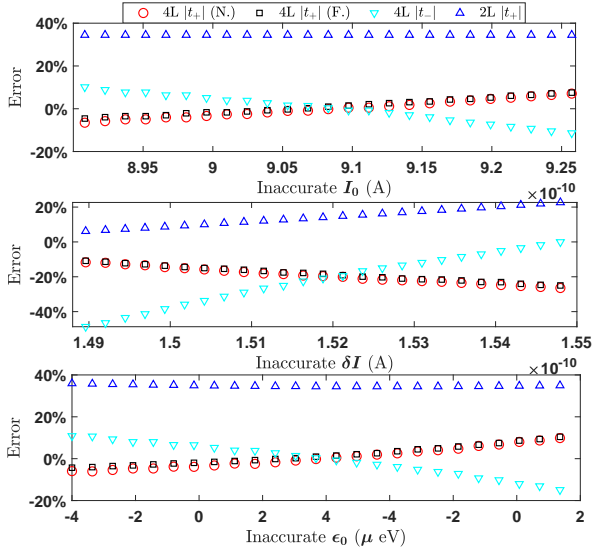} 
		\par\end{centering}
	\caption{\label{fig:guessI0dIep}Fitting error caused by wrong estimation of
		$I_{0}$, $\delta I$, and $\epsilon_{0}$. Parameters are the same as Table~\ref{tab:1}, QD~1-2.}
\end{figure}

In the ``Results'' section, we describe our fitting procedure by splitting the
fitting parameters into two groups: (1) $I_{0}$, $\delta I$, and
$\epsilon_{0}$, which determine the position of the fitting curve; and (2) $t_{+}$
and $t_{-}$, which determine the shape of the curve. In our protocol we fit the
two groups of parameters in turn until the results converge. Practically,
we only perform the iterative fitting 2 rounds because a bad estimate
on $I_{0}$, $\delta I$, and $\epsilon_{0}$ does not result in too
much error on the final fitting results of $t_{\pm}$. In Fig.~\ref{fig:guessI0dIep},
we plot the fitting error caused by a wrong estimate of $I_{0}$,
$\delta I$, and $\epsilon_{0}$. Fig.~\ref{fig:guessI0dIep} is
plotted in a relative wide range. However, practically
the errors cannot be too large, otherwise the best fitting curve will have
a notable shift. Therefore, the
iterative fitting converge very fast and 2 rounds of fitting is generally enough.

\subsection*{\label{Sec:App5}fitting error caused by incomplete knowledge of system parameters}

Valley parameters such as $|\Delta_{L}|$, $|\Delta_{R}|$ are crucial in describing a Si DQD.
As shown in Eq.~(\ref{eq:PLi}) and (\ref{eq:PL2}), the fitting protocol in our model requires a preliminary
measurement on the valley splitting of the two dots. Practically, the
valley splittings may be unknown or only roughly estimated. In the 
example in Table~\ref{tab:1}, the valley splittings are indeed estimated bu not measured. It is thus important to know the impact on the accuracy
of $t_{\pm}$ by inexact knowledge of $|\Delta_{L,R}|$. 

\begin{figure}
	\begin{centering}
		\includegraphics[width=1\columnwidth]{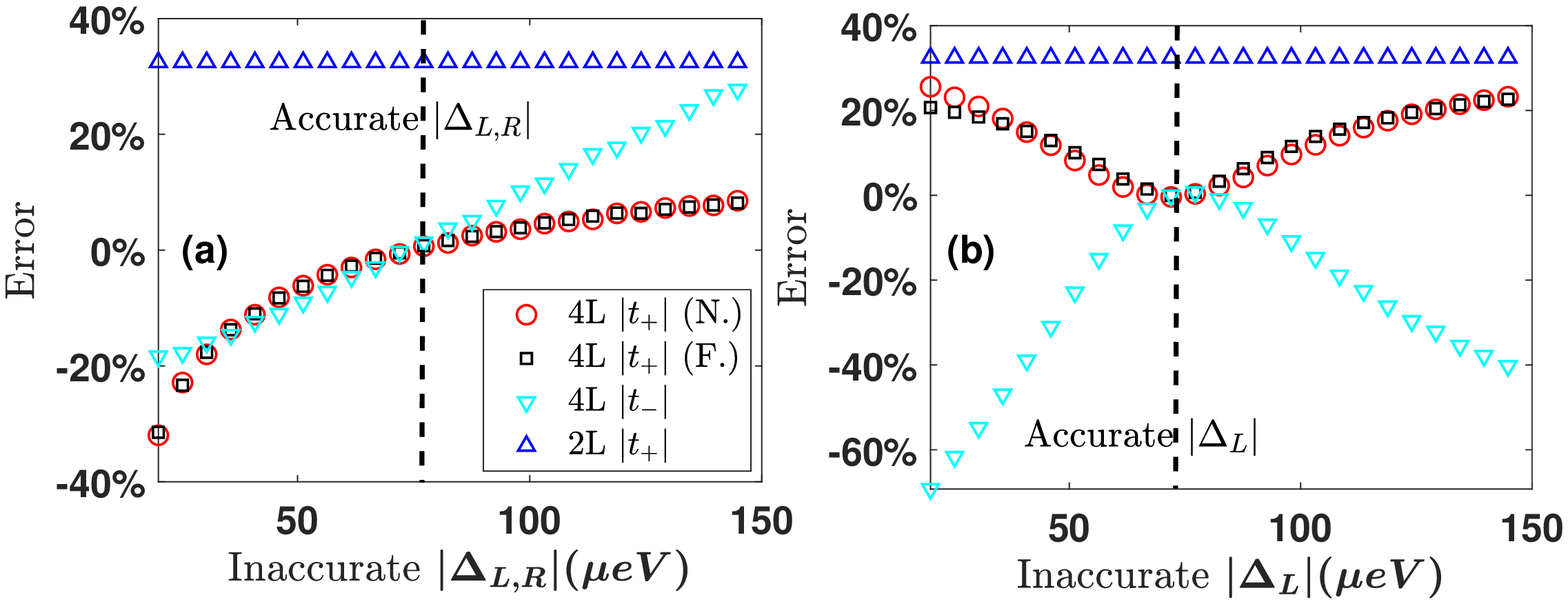}
		\par\end{centering}
	\caption{\label{fig:GuessVLR}Fitting error by different methods, the ``pseudo-curve''
		data is obtained with $t_{C}=0.071$ meV, $|\Delta_{L}|=|\Delta_{R}|=0.074$
		meV, and $\delta\phi=0.7\pi$. Inaccurate $|\Delta_{L,R}|$ indicate
		a wrong estimation/measurement on $|\Delta_{L,R}|$. In (a), we assume
		$|\Delta_{L}|=|\Delta_{R}|$. In (b), we assume $|\Delta_{R}|$ is
		measured accurately, only $|\Delta_{L}|$ has an error.}
\end{figure}

In Fig.~\ref{fig:GuessVLR}
we plot the error caused by incomoplete knowledge of valley splitting
$|\Delta_{L,R}|$. The numerical data used in fitting are generated
with $|\Delta_{L,R}|=74$ $\mu$eV. 
We then vary $|\Delta_{L}|$ or $|\Delta_{R}|$ on purpose to examine the sensitivity of our protocol to this systematic error. 

The numerical results show that a moderately off estimate of $|\Delta_{L,R}|$ will not lead to sizable errors in 4L fitting unless $|\Delta_{L,R}|$ is significantly underestimated. In Fig.~\ref{fig:GuessVLR}~(a), when $|\Delta_{L,R}|\approx37$ $\mu$eV (50\% underestimated), the 4L fitting error is still only about 10\%. Furthermore, an
overestimated $|\Delta_{L,R}|$ will result in even smaller error
comparing to an underestimated $|\Delta_{L,R}|$, and the 4L fitting
error is always smaller than the 2L fitting. The right panel (b) shows
the case that only one valley splitting ($|\Delta_{L}|$) is known
inaccurately. Similar to panel (a), the 4L fitting error is always smaller than
the 2L fitting. We note that the fitting accuracy of inter-valley tunnel
coupling \textbar$t_{-}$\textbar{} is more sensitive to the knowledge
of $|\Delta_{L}|$, though information on $t_-$ is not accessible at all to a 2L model, since valley physics is not included there.

Another source of error in tunnel couplings is the measurement of the current
$I_{PQC}$, which always contains some noise $\delta I_{noise}$ as
shown in Eq.~(\ref{eq:IQPC}) and illustrated in Fig.~\ref{fig:1}. Here, we use a stochastic
function $\delta I_{noise}$ to simulate the uncertainty in current
measurement. 
The stochastic function is characterized by the mean $\langle\delta I_{noise}\rangle=0$
and the standard deviation $\sigma(\delta I_{noise})$, which indicates
the strength of the noise. Numerical results as presented in Table~\ref{tab:3} show that the noise $\delta I_{noise}$
has a much larger impact on the performance of 2L fitting comparing
to the 4L fitting. By using the 4L model, even when the relative strength
of the noise reaches 5\%, the fitting error is still under 10\%, while
the 2L fitting always produces a significant error over 35\%. The errors from 4L fitting do increase rapidly with an increasing $\sigma(\delta I_{noise})$, while errors in the 2L fitting
remains large and do not change dramatically as $\sigma(\delta I_{noise})$ increases.
\begin{table}[h]
	\begin{tabular}{|c|c|c|c|c|c|}
		\hline
		$\frac{\sigma(\delta I_{noise})}{\delta I}$ & 0.01 & 0.02 & 0.03 & 0.04 & 0.05\tabularnewline
		\hline
		\hline
		Error on $|t_{+}|$ (2L) & 36.87\% & 36.53\% & 36.18\% & 36.82\% & 37.18\%\tabularnewline
		\hline
		Error on $|t_{+}|$ (4L) & 1.56\% & 4.30\% & 4.98\% & 6.25\% & 8.19\%\tabularnewline
		\hline
		Error on $|t_{-}|$ (4L) & 1.06\% & 2.81\% & 3.54\% & 4.16\% & 4.68\%\tabularnewline
		\hline
	\end{tabular}
	
	\caption{\label{tab:3}Fitting error caused by noise on $I_{QPC}$ measurement.}
	
\end{table}

\section*{DATA AVAILABILITY}
Data and an example of fitting code are available from the authors on reasonable request.

\section*{ACKNOWLEDGEMENTS}

This work is supported by the Army Research Office, Project No. W911NF1710257. The authors
thank Adam Mills and Jason Petta for providing original data and helpful discussions.

\section*{AUTHOR CONTRIBUTIONS}
X.Z. and X.H. contributed to theoretical/numerical/physical analysis and prepared the manuscript.

\section*{Additional Information}
Competing Interests: The authors declare no competing interests.

\bibliographystyle{apsrev4-1}
\bibliography{MeasureTc}

\end{document}